\documentclass[review]{elsarticle}

\usepackage{graphicx}
\usepackage{xspace}
\usepackage{hyperref} 
\biboptions{sort&compress}
\usepackage{amssymb}
\providecommand*{\ie}{\emph{i.\,e.}\xspace}%
\providecommand*{\et}{\emph{et\,al.}\xspace}%
\providecommand*{\tc}{$T_c$\xspace}%
\providecommand*{\fts}{Fe$_{1+y}$Te$_{1-x}$Se$_x$\xspace}%

\begin{document}

\begin{frontmatter}

\title{Magnetic neutron scattering studies on the Fe-based superconductor system \fts}

\author{Jinsheng Wen\corref{myead}}

\address{Center for Superconducting Physics and Materials,
National Laboratory of Solid State Microstructures, Department of Physics,
and Collaborative Innovation Center of Advanced Microstructures, Nanjing University, Nanjing 210093, China}

\cortext[myead]{jwen@nju.edu.cn}

\begin{abstract}
I present a brief overview on the interplay between magnetism and superconductivity in one of the Fe-based superconductor systems, \fts, where the research of our group has centered. The parent compound Fe$_{1+y}$Te is an antiferromagnet; with Se doping, antiferromagnetic order is suppressed, followed by the appearance of superconductivity; optimal superconductivity is achieved when $x\sim50\%$, with a superconducting temperature \tc of $\sim$15~K. The parent compound has an in-plane magnetic ordering wave vector around (0.5,\,0) (using the tetragonal notation with two Fe atoms per cell).  When Se concentration increases, the spectral weight appears to be shifted to the wave vector around (0.5,\,0.5), accompanying the optimization of superconductivity. A neutron-spin resonance has been observed around (0.5,\,0.5) below \tc, and is suppressed, along with superconductivity, by an external magnetic field. Taking these evidences into account, it is concluded that magnetism and superconductivity in this system couple to each other closely---while the static magnetic order around (0.5,\,0) competes with superconductivity, the spin excitations around (0.5,\,0.5) may be an important ingredient for it. I will also discuss the nature of magnetism and substitution effects of 3$d$ transition metals.   
\end{abstract}


\end{frontmatter}


\section{Introduction \label{sec:intro}}
\subsection{A brief overview}
The quest for an alternative superconducting mechanism was soon initiated after the discovery of high-temperature superconductivity in the lamellar copper-oxide materials~\cite{mullerlbco,ybcodis,ybcodis2}, which posed a great challenge to BCS theory~\cite{bcstheory}, the many-body theory developed by Bardeen, Cooper, and Schrieffer that successfully explained conventional superconductivity~\cite{bcstheory}. The key concept is that electrons form pairs aided by the electron-phonon interactions, and the pairs condense at the superconducting temperature \tc. In the cuprate superconductors, electron-phonon coupling is not sufficient to induce superconductivity with such high $T_c$s~\cite{lee:17,carlsonbook1}. In these materials, superconductivity develops from electronically doping a Mott insulator, and is in close proximity to the antiferromagnetic phase~\cite{lee:17,carlsonbook1,birgeneau-2006,RevModPhys.70.897,orenstein}. Thus, it is very promising that one may eventually work out the high-\tc mechanism by studying the interplay between magnetism and superconductivity.

\begin{figure}[ht]
\begin{center}
  \includegraphics[width=\linewidth]{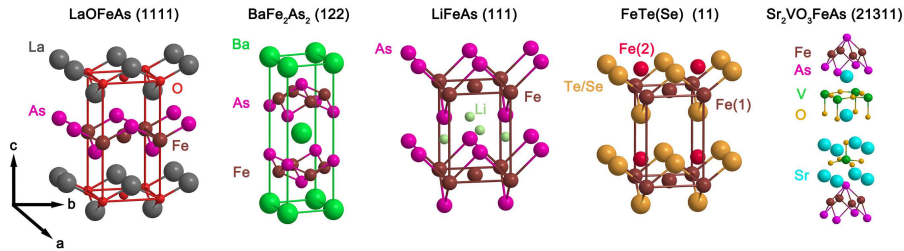}  \end{center}
  \caption{ Schematic crystal structures for the 1111, 122, 111, 11, and 21311 type Fe-based superconductors.}\label{fig:ironbasedsc}
\end{figure}

Research on this subject gained substantial momentum in 2008 with the discovery of superconductivity in compounds that contain Fe instead of Cu (termed ``Fe-based superconductors'')~\cite{hosono_1,hosono_2}. The field was initially excited by the discovery of superconductivity in LaFeAsO$_{1-x}$F$_x$ (labeled 1111 based on the elemental ratios in the chemical formula of the parent material) with $T_c=26$~K by Hosono's group~\cite{hosono_2}, following the group's earlier report of superconductivity in LaFePO$_{1-x}$F$_x$ with \tc $\sim$~5~K~\cite{hosono_1}. Soon after the initial discovery, the scientific community witnessed a burst of new Fe-based superconductors. So far, besides the 1111 system, other five major families of Fe-based superconductors have been discovered, typified by BaFe$_2$As$_2$ (122)~\cite{sefat:117004,rotter-2008-101,chen-2008-25,inosov:224503}, LiFeAs (111)~\cite{lifep,lifeas1,lifeas2,lifeas3,structure5,lifep2,nafeas1}, \fts (11)~\cite{hsu-2008,yeh-2008,sales:094521,chen:140509,fang-2008-78}, Sr$_2$VO$_3$FeAs (21311)~\cite{zhu:220512,sc21311}, and $A_{x}$Fe$_{2-y}$Se$_{2}$ ($A=$ alkaline elements)~\cite{scinkfese,kfese2,rbfese1,csfese1,tlrbfese1}. The crystal structures for the five families are shown in Fig.~\ref{fig:ironbasedsc} (The $A_{x}$Fe$_{2-y}$Se$_{2}$ compounds are isostructural to the 122)~\cite{structure5,zhao:132504,hosono_1,lester:144523,qiu:257002,kumar:144524}. They are all tetragonal at room temperature, and have layered structures. The current record of \tc in the bulk materials is 56~K in Gd$_{1-x}$Th$_x$FeAsO~\cite{epltc56k,ren-2008-25}. This makes the Fe-based superconductors second only to cuprates in \tc, and for this reason they are often considered as another class of high-temperature superconductors. Comparing the differences and similarities between the two classes may help find the common ground underlying the high \tc in these superconductors.

The distinct properties of the parent compound set the Fe-based superconductors apart from the copper oxides. The undoped cuprates are Mott insulators, which are predicted by band theory to be metallic but turn out to be insulating because the otherwise itinerant electrons are localized due to the large Coulomb repulsion~\cite{lee:17,carlsonbook1}. In the Fe-based superconductors, their parent compounds are metallic~\cite{ma:033111}. This naturally leads to a different starting point, and a weak-coupling theory is often more favorable~\cite{mazin:057003,kuroki-2008}. Furthermore, unlike cuprates where only a single Cu 3$d$ band is involved, four or five Fe 3$d$ orbitals are involved in the multiple bands that cross the Fermi level~\cite{kuroki-2008}. This in some cases can complicate the interpretations~\cite{PhysRevLett.104.157001}. Despite the differences, these two classes of high-temperature superconductors share surprisingly similar phase diagrams~\cite{mazinnaturereview}. With very few exceptions in the Fe-based superconductors, the parent compounds exhibit long-range antiferromagnetic order, which is suppressed with
doping, and superconductivity appears above a certain doping
level~\cite{cruz,lester:144523,qiu:257002,kumar:144524,huang:257003,chen:064515,johannes-2009,wilson:184519,kofu-2009-11,zhao-phasedgm,luetkens-2008,drew-2009,rotter-2008-47,chen-2009-85,fang:140508,chu:014506,khasanov:140511,liupi0topp,0295-5075-90-2-27011,spinglass,JPSJ.79.102001,revisedfetesephase}, resembling the phase diagrams of the cuprate superconductors~\cite{lee:17,carlsonbook1,birgeneau-2006,RevModPhys.70.897,orenstein}.  Such a similarity immediately leads to the speculation that the pairing mechanism may be the same, with magnetic excitations replacing phonons in the electron pairing interactions~\cite{mazin:057003,kuroki-2008,ma:033111,dong-2008-83,cvetkovic-2009,graser-2009,JPSJ.77.113703}.  However, because of the multiband nature of the superconductivity, orbital excitations have also been proposed as possible contributors to the pairing mechanism~\cite{PhysRevLett.104.157001,PhysRevLett.109.137001,Kontani2012718}. Although a consensus on the high-\tc mechanism has not been reached so far, it is generally believed that studying the magnetic correlations has and will continue to yield important results that are critical in understanding the high-\tc superconductivity. 

With the tremendous efforts on studying the Fe-based superconductors, there has been a plethora of literature on this topic. A great number of comprehensive and topical review articles are already available~\cite{2011arXiv1106.1618S,Norman08042011,2011arXiv1106.3712H,Wang08042011,interplaywen,np_8_709,JPSJ.78.062001,lynn-2009-469,lumsdenreview1,mkwreview1,canfieldreview,dcjohnstonreview,mazinnaturereview,JPSJ.79.102001,htcironbasereview}. In the current work, I will only focus on the \fts system, with emphasis on the neutron scattering studies of the magnetism within this system.

\subsection{\fts system}

Superconductivity in the \fts system was discovered in Fe$_{1+y}$Se, with zero resistance at $T\approx 8$~K~\cite{hsu-2008}. For this system, the highest \tc at ambient pressure was found to be $\sim$~15~K for Se concentration around 0.5~\cite{yeh-2008,fang-2008-78,sales:094521}. Under pressure, the \tc in Fe$_{1+y}$Se could be raised up to 37~K~\cite{medvedev-2009-8,margadonna:064506,JPSJ.78.063704}. A more significant enhancement of the \tc was reported by scanning tunneling microscopy (STM) measurements on a single-layer FeSe thin film~\cite{2012arxiv1201.5694W}.It was shown that there was a possible superconducting gap of 20~meV, suggesting that the \tc could be above 77~K, although transport measurements only found zero resistance below 30~K~\cite{2012arxiv1201.5694W}. Later angle-resolved-photoemission-spectroscopy (ARPES) measurements revealed a gap up to 20~meV that closed at $\sim$~65~K~\cite{nm_12_634,nm_12_605,arXiv:1202.5849}. More recently, transport measurements showed higher \tc than that observed in the initial report~\cite{2012arxiv1201.5694W,arXiv:1406.3435,sr4_6040}. In particular, superconducting-like electrical behavior was observed at temperatures above 100~K from the in-situ transport measurements on the monolayer FeSe films using a four-probe method~\cite{arXiv:1406.3435}. These studies suggested that 56~K, the record of $T_c$ for the Fe-based superconductors might have been broken~\cite{epltc56k,ren-2008-25,np10_892}.  

Aside from the possible high \tc in thin films, there are several other features that highlight the importance of this system. For example: i) The magnetism in its parent compound may have a different origin than in the other Fe-based compounds, as will be discussed in Sec.~\ref{sec:order}; ii) As can be seen from Fig.~\ref{fig:ironbasedsc}, the crystal structure of the 11 system is the simplest; iii) As is obvious from the chemical formula, the 11 system does not contain As, so it is safer to handle; iv) Last but not least, large-size single crystals can be made available for this system (with $x\leq0.7$). This is especially important for neutron scattering experiments because the limitations of neutron source strength and small scattering cross sections require the use of samples of considerable size ($\gtrsim 1$~cm$^3$) to obtain reliable data in the given beamtime.

\subsection{Neutron scattering experiments}
Before I proceed, I shall discuss neutron scattering, the most powerful tool in characterizing magnetic correlations. Neutron scattering is analogous to X-ray scattering~\cite{neutron8}. They both can be used to probe the crystal structure and lattice dynamics ($i.e.$ phonons), but since a neutron carries a spin of 1/2, it can interact with the spin of the unpaired electron, allowing the material's magnetic information to be revealed. Neutrons are charge neutral and interact with matter through short-range weak forces, thus they have large penetration depth. This makes the bulk properties easily accessible by neutrons. The following discussions will be mostly on the magnetic scattering, whose cross section can be written as~\cite{neutron1,neutron2,igorbook1}
\begin{equation}\label{eq:mdouble}
    \frac{d^2\sigma}{d\Omega dE_f}=\frac{N}{\hbar}\cdot\frac{k_f}{k_i}\cdot p^2\textnormal{e}^{-2W}S({\bf Q},E),
\end{equation}
where $N$ is the total number of the unit cells, $k_f$
and $k_i$ are the final and incident neutron wave vectors respectively, $p=\frac{\gamma r_{0}}{2}gf({\bf Q})$, $\frac{\gamma r_{0}}{2}=0.2695\times10^{-12}$~cm, $g$ is the Land\'{e} splitting factor, $f(\bf Q)$ is the
the wave vector {\bf Q}-dependent magnetic form factor, e$^{-2\rm W}$ is the Debye-Waller factor, and $S({\bf Q},E)$ is the {\bf Q}- and energy $E$-dependent dynamical spin correlation function, which is the quantity one would like to obtain eventually. $S({\bf Q},E)$ is related to the imaginary part of the dynamical spin susceptibility, $\chi''({\bf Q},E)$ via the fluctuation-dissipation theorem~\cite{gynormal},
\begin{equation}\label{eq:schi}
    S({\bf Q},E)=\frac{\hbar}{\pi g^2\mu_{\rm{ B}}^2}\cdot\frac{1}{1-\textnormal{e}^{-E/{\rm k_B}T}}\cdot\chi''({\bf Q},E), 
\end{equation}
where  $\mu_{\rm B}$ is the Bohr magneton, and $(1-\textnormal{e}^{-E/{\rm k_B}T})^{-1}$ is referred to as the Bose factor. It also satisfies a simple moment sum rule for a system with spins $S$ when one performs the integration over a Brillouin zone, $\int_{\rm BZ} S({\bf Q},E)d^3{\bf Q}dE=S(S+1)$~\cite{neutron1,neutron2}.

Triple-axis spectrometers (TAS) and time-of-flight spectrometers (TOFS) that can probe both elastic and inelastic scattering processes are commonly used. For a TAS, the three axes correspond to monochromator, sample, and analyzer axes. When doing an experiment with a TAS, one often fixes the final energy $E_f$ and varies the incident energy $E_i$. The energy is usually determined from the neutron wave length $\lambda$ by $E=81.79/\lambda^2$, and $\lambda$ is determined using the reflection from a single-crystal plane according to the Bragg law $\lambda=2d\rm{sin}\theta$. The spacing of the reflection plane $d$ is defined by $d=2\pi/\sqrt{\frac{H^2}{a^2}+\frac{K^2}{b^2}+\frac{L^2}{c^2}}$, with $(HKL)$ being the Miller index, and $a$, $b$, $c$ being the lattice constants. The angle between the incident beam and the plane is $\theta$. Pyrographites with (002) plane are often used as the monochromator and analyzer. It is quite convenient to probe $S({\bf Q},E)$ at a specific $\bf Q$-$E$ point precisely utilizing a TAS, with good ${\bf Q}$ and $E$ resolution. It is suitable to measure systems where the signals are concentrated at certain regions. The disadvantage is that it only accesses a single point of the momentum-energy space at a time. If the spectral weight spreads over a wide range of $\bf Q$ and $E$, it will be very time consuming doing such an experiment. Also, the energy range is limited. For a TOFS, there is an array of position-sensitive detectors that allow a portion instead of a point of the momentum-energy space to be covered. The energies in this case are measured by the time neutrons taken to travel. Compared with the TAS, the availability of the TOFS is more limited, and the resolution is often more coarse for certain type of TOFS. Since they both have merits of their own, they are often used in complement to each other.

In describing the neutron scattering data, the wave vector $\bf Q$ of ({\bf Q}$_x$,{\bf Q}$_y$,{\bf Q}$_z$) expressed in $(HKL)$ has the reciprocal lattice unit (rlu) of $(a^*,b^*,c^*)=(2\pi/a,2\pi/b,2\pi/c)$. Throughout this paper, the tetragonal notation with two Fe atoms per cell is used (as shown in Fig.~\ref{fig:magneticstructure}). In this case the lattice constant $a\approx 3.8$~\AA.

\subsection{Organization of the article}
The remainder of this paper is organized as follows: in Sec.~\ref{sec:order}, results on the static magnetic order in \fts will be presented; in Sec.~\ref{sec:spindynamics}, the spin dynamics of this system will be shown; in Sec.~\ref{sec:nature}, the origin of the magnetic excitations will be discussed briefly; in Sec.~\ref{sec:subeffect}, the substitution effects using 3$d$ transition metals will be summarized, followed by the conclusions in Sec.~\ref{sec:conclusion}.

\section{Static magnetic order \label{sec:order}}

\begin{figure}[ht]
\begin{center}  \includegraphics[width=0.8\linewidth]{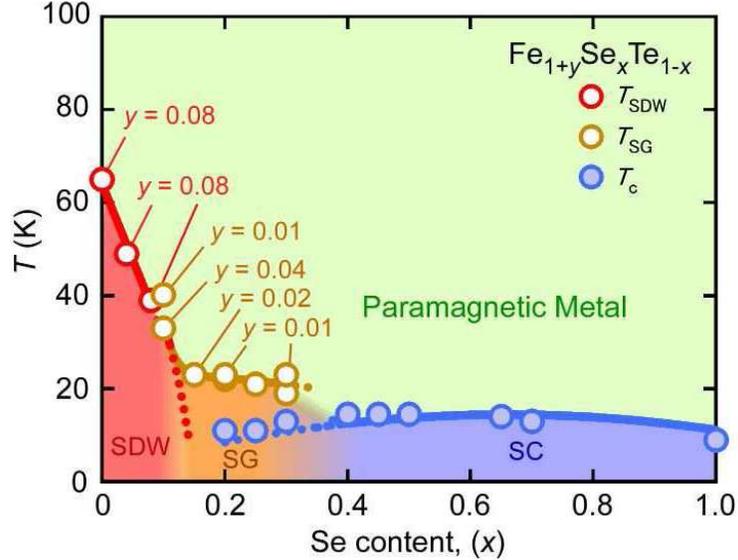}
\end{center}  \caption{Phase diagram of \fts, constructed from resistivity, magnetization and neutron scattering data on single crystals. The nominal Fe content is $y=0$, unless specified. SDW, SG, and SC stand for antiferromagnetic, spin-glass and superconducting phases, respectively. Reprinted from~\cite{spinglass}. $\copyright$ 2010 Physical Society of Japan.}\label{fig:fetesephasedgm}
\end{figure}

It is convenient to discuss the magnetic order with reference to the phase diagram as shown in Fig.~\ref{fig:fetesephasedgm}~(Ref.~\cite{spinglass}). It was obtained by performing resistivity, magnetic susceptibility and neutron scattering measurements on single crystals grown using the Bridgman method~\cite{interplaywen}. Some modified versions have also been made by several other groups~\cite{khasanov:140511,liupi0topp,0295-5075-90-2-27011,spinglass,JPSJ.79.102001,revisedfetesephase}, but the basic features are similar to those shown in Fig.~\ref{fig:fetesephasedgm}. At a first glance, this phase diagram is similar to those of the cuprate~\cite{lee:17,birgeneau-2006,RevModPhys.70.897,orenstein,carlsonbook1}
and iron-pnictide superconductors~\cite{cruz,qiu:257002,huang:257003,chen:064515,johannes-2009,wilson:184519,kofu-2009-11,zhao-phasedgm,luetkens-2008,drew-2009,rotter-2008-47,chen-2009-85,fang:140508,chu:014506}. 
There are several important differences which need to be pointed out though: i) Unlike most other high-\tc superconductors, where ``doping" often refers to substitution of an element with another that has a different valence, here Te is isovalent to Se; ii) Superconductivity survives to $x=1$, unlike most other systems where superconductivity disappears with carrier doping above a certain value~\cite{lumsdenreview1}; iii) The material's properties can be tuned not only by doping with Se, but also by adjusting the amount of excess Fe, so when writing the chemical formula for this system, there is a second variable. Indeed, the extra Fe is required for samples with small $x$ to stabilize the crystal structure.

\begin{figure}[ht]
\begin{center}
  \includegraphics[width=0.9\linewidth]{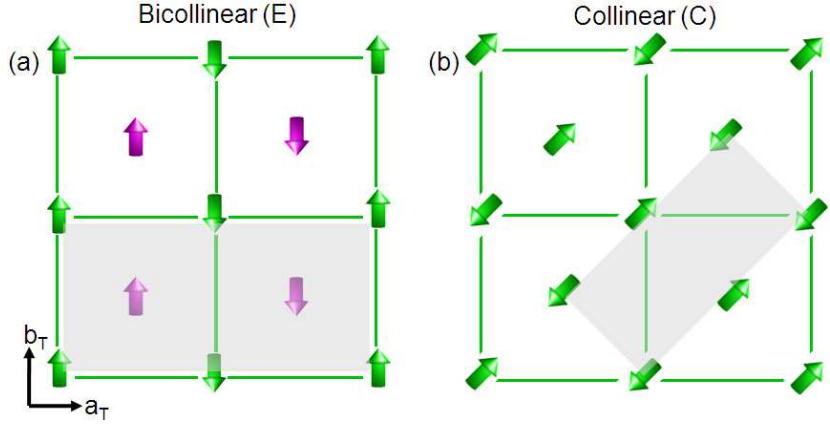}\end{center}
  \caption{(a) Schematic for the bicollinear (E-type) in-plane spin structure in the tetragonal notation of the 11 compound. (b) Collinear (C-type) magnetic structure for iron pnictides. Shadow represents the magnetic unit cell.}\label{fig:magneticstructure}
\end{figure}

Fe$_{1+y}$Te is not superconducting; it exhibits coincident magnetic and structural transitions at  $\sim65$~K~\cite{bao-2009,li-2009-79,liupi0topp,PhysRevB.81.094115}. This behaviour is similar to that of the undoped phases of the 1111 and 122 materials~\cite{cruz,qiu:257002,huang:257003,chen:064515,johannes-2009,wilson:184519,kofu-2009-11,zhao-phasedgm,luetkens-2008,drew-2009,rotter-2008-47,chen-2009-85,fang:140508,chu:014506,khasanov:140511,0295-5075-90-2-27011,JPSJ.79.102001}, and it is therefore often referred to as the parent compound for the 11 system~\cite{bao-2009,li-2009-79,mizuguchi-2009-94,mizuguchi-2009-469,fang-2008-78}. Initial band-structure calculations predicted that its Fermi-surface topology should be similar to that of the iron pnictides~\cite{subedi-2008-78}; this had been confirmed by ARPES measurements~\cite{xia037002,PhysRevLett.105.197001}. From the Fermi-surface-nesting picture~\cite{subedi-2008-78}, one would expect that this system would have a collinear (C-type) spin-density-wave (SDW) order with an in-plane wave vector (0.5,\,0.5), as shown in Fig.~\ref{fig:magneticstructure}(b). However, this is in contrast to the experimental results showing a magnetic ordering vector of (0.5,\,0) by Fruchart \et \cite{structure6} in Fe$_{1.125}$Te a few decades ago. This result was confirmed by Bao \et~\cite{bao-2009} in Fe$_{1.075}$Te, and by Li \et~\cite{li-2009-79} in Fe$_{1.068}$Te; each of these had a bicollinear (E-type) spin structure as shown in Fig.~\ref{fig:magneticstructure}(a). Even more surprising is the fact that ARPES measurements had observed no SDW nesting instability along [0.5,\,0]~\cite{xia037002,Physics.2.59}. Clearly, a simple nesting mechanism cannot account for these experimental results.

The magnetic order in Fe$_{1+y}$Te is long ranged, with a moment size of $\sim2.5\ \mu_{\rm B}$/Fe for $y=0.05$ at the base temperature~\cite{PhysRevB.81.094115}.  The moment size is significant compared to that in the iron-pnictide antiferromagnets (less than 1~$\mu_{\rm B}$/Fe)~\cite{wilson:184519,cruz}, but it is small compared to the effective moment estimated from the magnetic susceptibility in the paramagnetic phase~\cite{2011arXiv1103.5073Z}. The moment is found to be aligned mostly along the $b$ axis as shown in Fig.~\ref{fig:magneticstructure}(a)~\cite{li-2009-79,bao-2009,PhysRevB.81.094115}. The large moment together with the absence of the Fermi-surface-nesting feature along the ordering wave vector suggests that the magnetic order should be different from that of the Fe-pnictides where itinerant electrons dominate the magnetism. First-principle calculations considering the role of local moments and the importance of Hund's exchange coupling have provided better agreement with the experimental observations~\cite{han:067001,ma-2009,loalm1,johannes-2009,fang-2009,moon057003,weiguounified}. 

The order also depends on the amount of interstitial iron~\cite{bao-2009,2011arXiv1111.4236P,2012arXiv1202.4152S,PhysRevLett.112.187202}. The excess Fe resides in an interstitial site centered above (or below) a plaquette of four Fe sites~\cite{PhysRevB.81.094115,prbliudiffuse}. The magnetic interaction between an interstitial and the Fe nearest-neighbors results in locally ferromagnetic correlations~\cite{2011arXiv1103.5073Z,2011arXiv1109.5196T} and weak localization of the charge carriers~\cite{zhang:012506}. Rodriguez \et have constructed a magnetic-crystallographic phase diagram of Fe$_{1+y}$Te~\cite{PhysRevB.84.064403}. They have found that the order changes as $y$ increases through the critical excess Fe content, $y_c$, from bicollinear commensurate to helical incommensurate ~\cite{bao-2009,li-2009-79,PhysRevB.84.064403,2011arXiv1108.5968Z,2011arXiv1111.4236P}.  The change in magnetic order correlates with the change in the low-temperature structural symmetry from monoclinic to orthorhombic~\cite{PhysRevB.84.064403,arXiv:1202.2484}. A very recent theoretical calculation explains this evolution as the excess Fe modifying the local moments' Ruderman-Kittel-Kasuya-Yosida (RKKY) interaction, which depends on the Hund's coupling between the itinerant electrons and localized moments that is changing with $y$~\cite{arXiv:1408.1418}. It is worth mentioning that attempts utilizing spin-polarized STM to provide real-space imaging of the magnetic order in Fe$_{1+y}$Te have also been made, and the results appear to be consistent with those obtained from neutron diffraction measurements~\cite{Enayat08082014}. 

Upon doping with Se, the order is suppressed, with a reduced ordering temperature (from $65$~K to $\lesssim30$~K with 0.1 Se)~\cite{spinglass}, reduced size of the ordered magnetic moment (from 2.1~$\mu_{\rm B}$/Fe to 0.27~$\mu_{\rm B}$/Fe)~\cite{liupi0topp}, and shorter correlation length (the magnetic peak is resolution limited in the parent compound, while with 0.25 Se doping, the order is short ranged with a correlation length of $\sim$~4~\AA~\cite{wen:104506,bao-2009}). For Se content above 0.15, there is a coexistence between the spin-glass and superconducting phase~\cite{wen:104506,spinglass,khasanov:140511,liupi0topp,PhysRevB.81.094115}.   Interestingly, a magnetic peak is only observed on one side of the commensurate wave vector (0.5,\,0), [\ie (0.5-$\delta$,\,0) and not (0.5$\pm\delta$,\,0) with $\delta$ being the incommensurability]; this is likely a result of an imbalance of ferromagnetic/antiferromagnetic correlations between neighbouring spins~\cite{wen:104506}. With further increase of the Se concentration, the superconductivity optimizes with $x\approx0.5$, and no static magnetic order is observed (for $y\sim0$)~\cite{fieldeffectresonancewen,khasanov:140511,liupi0topp}.

Excess Fe atoms also affect the magnetic correlations in the samples with Se doping. Xu \et~\cite{xudoping11} have measured four samples: FeTe$_{0.7}$Se$_{0.3}$, Fe$_{1.05}$Te$_{0.7}$Se$_{0.3}$; FeTe$_{0.5}$Se$_{0.5}$, and Fe$_{1.05}$Te$_{0.55}$Se$_{0.45}$. Both samples with $y=0$ are superconducting with the same \tc of $\sim$~15~K, although the one with $x=0.3$ has a smaller superconducting volume fraction. Also, there is short-range static magnetic order near (0.5,\,0) in FeTe$_{0.7}$Se$_{0.3}$, while the sample with $x=0.5$ does not exhibit magnetic order, short or long ranged. With 0.05 extra Fe, superconductivity in both samples is fully suppressed, and the magnetic spectral weight around (0.5,\,0) is strongly enhanced. In both samples, there is short-range static order. These results clearly show that the excess Fe stabilizes the antiferromagnetic order but anticorrelates with the superconductivity. 

Bendele and coworkers have constructed a three-dimensional phase diagram with two variables, concentrations of Se, $x$, and excess Fe, $y$~\cite{PhysRevB.82.212504}. The dependence of superconductivity and magnetic order on both $x$ and $y$ can be more clearly seen from that phase diagram [Fig.~3(b) of Ref.~\cite{PhysRevB.82.212504}].

Since it is very challenging to grow single crystals of \fts with $x\gtrsim70\%$ because of the large vapor pressure of Se, neutron scattering results on the Se-rich side have been very limited~\cite{interplaywen,lumsdenreview1}. In the end member Fe$_{1+y}$Se, although it exhibits a symmetry-lowering structural transition on cooling through $\sim90$~K~\cite{fesephasetr,B813076K,PhysRevB.80.024517}, measurements with local probes such as M\"ossbauer spectroscopy~\cite{fesephasetr,mcqueen:014522} and $^{77}$Se nuclear-magnetic-resonance (NMR)~\cite{imai-2009-102} indicate absence of static magnetic order. In studies of a sample that shows an increase of $T_c$ to 37~K under pressure, M\"ossbauer measurements indicate absence of  magnetic order up to $\sim30$~GPa~\cite{medvedev-2009-8}. In contrast, Bendele \et~\cite{PhysRevLett.104.087003} have reported evidence for short-range magnetic order for $P\gtrsim0.8$~GPa based on muon-spin rotation ($\mu$SR) measurements; however, this sample shows a much more modest impact of pressure on the superconductivity, with a maximum \tc of 13~K at 0.7~GPa. ARPES measurements by two different groups on the thin-film samples of FeSe have detected features suggesting the existence of magnetic order for certain range of carrier dopings~\cite{nm_12_634,nm_12_605}. First-principle calculations have shown that the order can be of collinear type, the same as that in the Fe-pnictides~\cite{PhysRevB.89.014501}, or a unique block-checkerboard pattern~\cite{arXiv:1407.7145}.

\section{Spin dynamics \label{sec:spindynamics}}
\subsection{Spin dispersions \label{subsec:dispersion}}
Spin excitations in the parent compound Fe$_{1+y}$Te which orders antiferromagnetically below the N\'{e}el temperature $T_N$ have been first reported by two groups~\cite{PhysRevLett.106.057004,2011arXiv1103.5073Z}. Lipscombe \et have shown that their data can be fitted in
terms of spin waves using the Heisenberg model, with anisotropic nearest-neighbour (NN) and isotropic next-nearest-neighbour (NNN) couplings~\cite{PhysRevLett.106.057004}. In the work by Zaliznyak \et, it is found that a model with plaquettes of four ferromagnetically coaligned nearest-neighbour Fe spins that emerge as a new collective degree of freedom and have short-range antiferromagnetic correlations between the neighbouring plaquettes can more accurately describe their data~\cite{2011arXiv1103.5073Z}. 

\begin{figure}[ht]
\begin{center}
  \includegraphics[width=\linewidth]{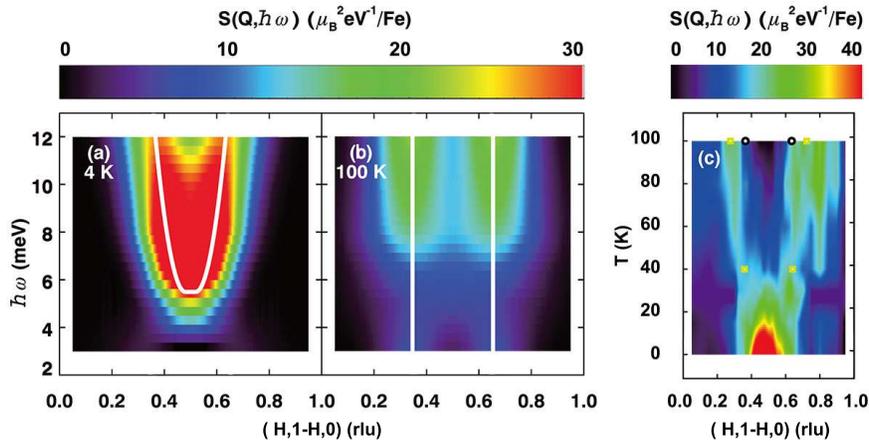}\end{center}
  \caption{(a) and (b), the magnetic spectrum of a superconducting FeTe$_{0.35}$Se$_{0.65}$ sample at temperature below and well above the \tc of 14~K; (c), temperature dependence of the magnetic scattering at $E=\hbar\omega=5$~meV. Data are extracted from Refs.~\cite{xulocal1,PhysRevLett.109.227002}.}\label{fig:spectrum}
\end{figure}

There are many more reports on the spin excitations in the Se-doped materials which exhibit no long-range antiferromagnetic order. In the optimally-doped samples, the spin excitations emerge from the in-plane wave vector of (0.5,\,0.5), different from that of the parent compound~\cite{lumsden-2009,shcouplingprb}. The evolution of the magnetic scattering from (0.5,\,0) to (0.5,\,0.5) will be discussed in Sec.~\ref{subsec:weighttransfer}. The spin-excitation spectrum of a superconducting FeTe$_{0.35}$Se$_{0.65}$ sample is shown in Fig.~\ref{fig:spectrum}~(Refs.~\cite{xulocal1,PhysRevLett.109.227002}). The sample has a \tc of 14~K. In the superconducting phase [Fig.~\ref{fig:spectrum}(a)], there is very little
spectral weight below 5~meV, indicating the opening of a spin-excitation gap of 5$\pm1$~meV. The magnetic scattering appears to be commensurate for energies around 5~meV, and becomes incommensurate at higher energies, thus displaying an upward parabola shape. A number of groups have reported similar spectra~\cite{lumsden-2009,shcouplingprb,PhysRevLett.105.157002,incomfetese1}.
In fact, the dispersion is anisotropic, with the scattering pattern elongating only along the [1$\bar{1}0$] direction in the $(HK0)$ plane, whereas the crystal has a four-fold symmetry~\cite{shcouplingprb,interplaywen}. Lee and coworkers have shown that by taking into account the spin-orbital coupling effects, this anisotropy can be explained~\cite{shcouplingprb}. For high temperatures, the shape of the spectrum changes, with the scattering being clearly incommensurate at the whole energy range, and the spectrum looks like a ``waterfall", as shown in Fig.~\ref{fig:spectrum}(b)~\cite{xulocal1}.

For samples with Se doping but without superconductivity ($e.g.$, with more excess iron), the scattering is incommensurate, similar to that of the superconducting samples at high temperatures~\cite{fetesecomp142202,PhysRevB.89.174517}. This indicates that the incommensurate to commensurate transition of the low-energy magnetic scattering in the superconducting samples may be related to superconductivity, but intriguingly, it is found that this transition occurs at a temperature significantly higher than \tc~\cite{PhysRevLett.109.227002,PhysRevB.89.174517}. By following the temperature dependence of the scattering at 5~meV in the superconducting samples, which is centered at (0.5,\,0.5) at low temperatures, it is shown that it splits into two incommensurate peaks displaced from (0.5,\,0.5) at $\sim$~40~K, as shown in Fig.~\ref{fig:spectrum}(c)~\cite{PhysRevLett.109.227002}. Theoretically, it has been proposed that short-range ordering of the Fe $d_{xz}$ and $d_{yz}$ orbitals will promote the Fermi-surface nesting between the electron and hole pockets~\cite{PhysRevB.80.064517,PhysRevB.85.024534}. Such a Fermi-surface topology can give rise to the commensurate scattering based on the Fermi-surface-nesting picture~\cite{subedi-2008-78}. Under the current theoretical framework, a nested Fermi surface is necessary for superconductivity, provided that the nesting is not strong enough to induce long-range magnetic order~\cite{PhysRevB.81.054502,PhysRevB.80.140515,cvetkovic-2009,mazin:057003,kuroki-2008}.
The short-range orbital ordering disappears with increasing disorder or temperature, resulting in the incommensurate excitations as well as the suppression of superconducting correlations. Lee \et have done detailed random phase approximation (RPA) calculations, and identified the role of orbital correlations~\cite{PhysRevB.86.094516}. The calculated imaginary part of the spin susceptibility agrees well with the experimental data.~\cite{PhysRevB.86.094516}

\subsection{The resonance mode \label{sec:resonance}}
\begin{figure}
\centering\includegraphics[width=\linewidth]{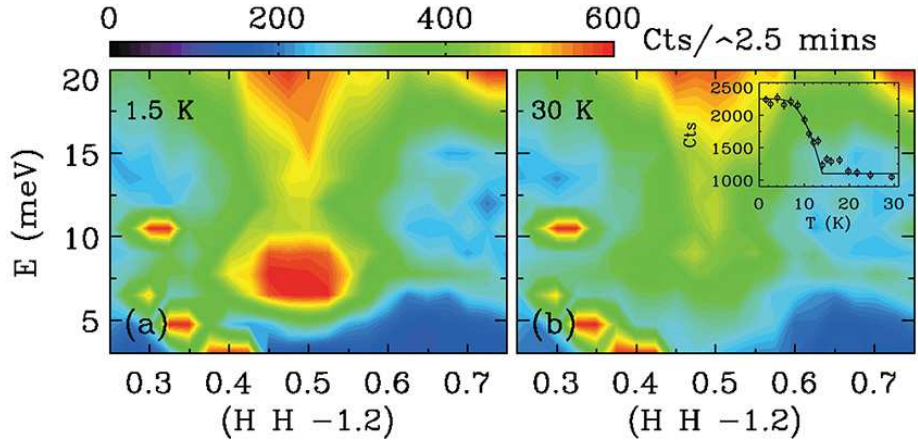}
\caption{ Spin resonance in FeTe$_{0.6}$Se$_{0.4}$. (a) and (b) show the spin excitation spectrum as a function of $\bf Q$ and energy at 1.5 and 30~K respectively. Inset in (b) shows the resonance intensity as a function of temperature. Data are obtained from Ref.~\cite{qiu:067008}.}\label{fig:feteseresonance}
\end{figure}

In this section, I will discuss the ``resonance" mode in the magnetic excitations, which is probably the most prominent feature of unconventional superconductivity~\cite{yu-2009np}. It describes the observation in inelastic neutron scattering experiments that the magnetic scattering at a particular $E$ and {\bf Q} has a significant enhancement when the system enters the superconducting phase. The resonance may provide important information relevant to the pairing symmetry, as it is predicted to occur at a particular wave vector that connects portions of the Fermi surface having opposite signs of the superconducting gap function~\cite{maier:020514,maier:134520,PhysRevLett.75.4126,PhysRevB.64.172508}.  Resonance excitations have been observed in a number of Fe-based superconductors~\cite{christianson-2008-456,resonance1111}, consistent with the presumed gap-sign change between the hole and electron pockets~\cite{mazin-2009}. In \fts, the resonance excitation has been observed to be near the (0.5,\,0.5) wave vector by several groups~\cite{qiu:067008,mook-2009-2,incomfetese1,shcouplingprb,fieldeffectresonancewen,PhysRevLett.105.157002,fetesecomp142202}.
Note that this wave vector is rotated by 45$^\circ$ from the magnetic ordering wave vector, whereas for iron pnictides, both magnetic order and resonance occur around the same wave vector of (0.5,\,0.5)~\cite{christianson-2008-456,resonance1111}. In fact, it is incommensurate in {\bf Q}, peaking at ($0.5\pm\delta$,$0.5\mp\delta$) in a direction transverse to (0.5,\,0.5)~\cite{shcouplingprb}. Compared to the magnetic excitation spectrum above \tc~[Fig.~\ref{fig:feteseresonance}(b)], the low-temperature spectral weight is greatly enhanced around (0.5,\,0.5) and the resonance energy of $\sim$~6.5~meV, as shown in Fig.~\ref{fig:feteseresonance}(a)~\cite{qiu:067008}. The resonance is a quasi two-dimensional excitation, with little $L$ dependence~\cite{lumsdenreview1}. The resonance energy corresponds to $\sim 5{\rm k_B}T_c$, similar to those of other high-\tc superconductors~\cite{lumsdenreview1}.  Accompanying the resonance, there is a spin gap with an energy of 5$\pm1$~meV; the intensity below the spin gap is shifted to the resonance in the superconducting state~\cite{qiu:067008,fieldeffectresonancewen}. The temperature dependence of the resonance intensity looks like the superconducting order parameter, as shown in the inset of Fig.~\ref{fig:feteseresonance}(b), demonstrating its close relationship with superconductivity. 

Since superconductivity can be suppressed by an external magnetic field, one will
naturally expect that the field should impact the resonance if it is related to the pairing, as seen in YBa$_2$Cu$_3$O$_{6.6}$~\cite{nature965} and in
La$_{1.82}$Sr$_{0.18}$CuO$_4$~\cite{tranquada-2004}. Experiment on FeTe$_{0.5}$Se$_{0.5}$ indeed gives positive answers~\cite{fieldeffectresonancewen}. It is shown that under an external field of 7~T, the resonance starts to appear at a lowered \tc, 12~K, with reduced intensity, coincident with the suppression of the superconductivity; however, there is no detectable change in either the resonance energy or the width of the resonance peak~\cite{fieldeffectresonancewen}. With a field of 14.5~T, Zhao \et~\cite{zhao-2010field122} have shown that in BaFe$_{1.9}$Ni$_{0.1}$As$_2$, both the resonance energy and intensity are reduced, and the resonance peak is slightly broadened.

It is believed by many researchers that the spin resonance is a singlet-to-triplet excitation of the Cooper pairs~\cite{PhysRevB.64.172508,PhysRevLett.75.4126,Bourges200545}. In principle, inelastic neutron scattering experiments under magnetic field should be able to test this hypothesis.  A magnetic field should induce a Zeeman splitting and lift the degeneracy of the triplet excited state~\cite{nature965}. Bao \et~\cite{wbao2} have tried to address this problem by applying a 14-T field on FeSe$_{0.4}$Te$_{0.6}$, and it appears that the field induces two additional peaks, consistent with the assumption that the resonance is a triplet. However, a more recent experiment shows that the field only reduces the spectral weight around the resonance mode~\cite{2011arXiv1105.4923L}. Perhaps relevant to this debate is a recent magnetic-field study on the resonance in the heavy-fermion superconductor CeCoIn$_5$, which reports that the resonance is a doublet, as the field splits it into two peaks~\cite{PhysRevLett.109.167207}. This result is however in contrast to the earlier studies where only reduction of the resonance energy is found~\cite{JPSJS.80SB.SB023,JPSJ.78.113706}. With no doubts, the debates on the nature of the resonance mode will be continued.

\subsection{Doping evolution of the magnetic excitations}
\label{subsec:weighttransfer}
\begin{figure}[ht]
\begin{center}\includegraphics[width=0.9\linewidth]{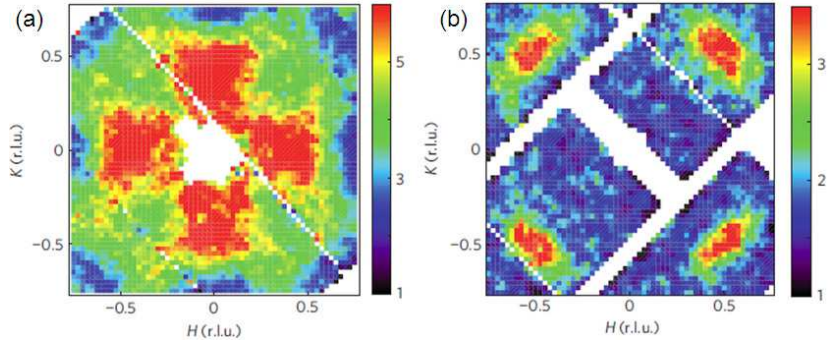}
\end{center}\caption{(a) Constant-energy cut of the magnetic excitation spectrum at an energy transfer of $6\pm1$~meV for the non-superconducting sample Fe$_{1.04}$Te$_{0.73}$Se$_{0.27}$ at 5~K. (b) Plot for the superconducting sample FeTe$_{0.51}$Se$_{0.49}$ at 3.5~K. Reprinted with permission from \cite{lumsden-2009}. $\copyright$ 2010 Macmillan Publishers Ltd.}
\label{fig:weighttransfer}
\end{figure}

As shown in Sec.~\ref{sec:order}, the non-superconducting parent compound Fe$_{1+y}$Te has static magnetic order with a bicollinear spin configuration. In these samples, spin excitations emerge from (0.5,\,0). In samples with robust superconducting properties, there is strong magnetic scattering around (0.5,\,0.5) with a spin resonance below \tc. The excitations correspond to spin correlations of the collinear type. It is natural to ask how the spin dynamics evolve from the parent compound to the optimally-doped materials. A number of theoretical~\cite{weiguounified,fang-2009} and experimental~\cite{lumsden-2009,liupi0topp,xudoping11,fetesecomp142202} studies have been carried out on the doping evolution of the magnetic correlations (static and dynamic) and their connection with superconductivity. Lumsden \et~\cite{lumsden-2009} have performed measurements using TOFS on two samples, a non-superconducting Fe$_{1.04}$Te$_{0.73}$Se$_{0.27}$ sample and a superconducting FeTe$_{0.51}$Se$_{0.49}$ sample. It is clearly shown in Fig.~\ref{fig:weighttransfer}(a) that at 5-7~meV, for the non-superconducting sample, the spectral weight is mostly concentrated around (0.5,\,0), where static magnetic order is observed. For the superconducting sample, magnetic excitations near (0.5,\,0.5) are dominant, as shown in Fig.~\ref{fig:weighttransfer}(b). On the other hand, the high-energy (larger than 120~meV) spectrum looks qualitatively similar for these two samples~\cite{lumsden-2009}. Several other groups have shown similar results that support the same conclusions: the spectral weight around (0.5,\,0) is related to the static magnetic order that competes with superconductivity, while the collinear spin correlations appear to be strengthened as superconductivity is optimized~\cite{liupi0topp,xudoping11,fetesecomp142202,PhysRevB.89.174517}.

The second variable $y$ in \fts also affects the magnetic excitations. By systematically studying the effects of excess Fe, it is shown in Ref.~\cite{xudoping11} that increasing excess Fe acts like reducing the amount of Se in terms of tuning the spin excitations---with more Fe, the low-energy excitations close to (0.5,\,0) are enhanced, while those close to (0.5,\,0.5) are suppressed, along with superconductivity. The extra Fe can induce short-range static magnetic order in the samples where no such order is present. These results reinforce the idea that the bicollinear E-type spin correlations are harmful for superconductivity, while the collinear C-type correlations may be compatible with it. In the parent compound Fe$_{1+y}$Te, Stock \et~\cite{2011arXiv1103.1811} have also shown that by changing $y$, the low-energy magnetic excitation spectrum can be dramatically modified, thus demonstrating the important role of excess Fe.

\section{Nature of the magnetism \label{sec:nature}}
As in the cuprates, the origin of the magnetism in the Fe-based compounds is heatedly debated~\cite{guangyongbi2212}. There are three main scenarios on this debate. First, there is the localized-moment picture, where strong correlation and Hund's coupling dominate~\cite{loalm1,localm2,localm3}. Second, there is the opposite case, where itinerant electrons contribute to the moments at low temperature as a result of the SDW instability, which depends critically on the nesting properties of the Fermi surface~\cite{cvetkovic-2009,singh:237003,diallo:187206,yin:047001,mazin:057003,han:067001,matan:054526,subedi-2008-78,zhang:012506,2010stoner1}. Given the metallic behavior of the Fe-pnictides, this idea is generally more favorable. A third possibility is intermediate between these two
cases, where some of the $d$-electrons are indeed itinerant, while others are localized, giving rise to an intermediate-size moment~\cite{hybrid1,arXiv:1202.2827,np_8_709,arXiv:1408.1418}. In addition to these three cases, there is the possibility that none of these is correct, but instead, that the magnetism derives simply from the energy gain associated with transfer to lower energy of one-electron density-of-states (DOS) spectral weight~\cite{johannes-2009}.

For \fts, opinions are more divided compared with other Fe-based superconductors~\cite{Physics.2.60,Physics.2.59}, as it has stronger electronic correlation~\cite{electroniccorre}, larger magnetic moment~\cite{bao-2009}, and poorer metallic properties~\cite{interplaywen}. More importantly, as discussed in Sec.~\ref{sec:order}, a simple nesting picture in an itinerant electron model cannot describe the magnetic order~\cite{xia037002}. However, for the magnetic excitations around (0.5,\,0.5), Argyriou \et have shown that the incommensurate spin excitation spectrum can be fitted well assuming that the excitations are quasiparticle scatterings between the hole and electron bands~\cite{incomfetese1}, but the lack of temperature dependence of the spectral weight and the large moment size do not agree with the itinerant model~\cite{xulocal1}. X-ray emission spectroscopy measurements on \fts suggest that the magnetism has a dual nature, with contributions from both localized spins and itinerant electrons~\cite{PhysRevB.84.100509}. This is consistent with the transport and magnetic susceptibility results that show both itinerant electrons and local moments exist in the system~\cite{chen:140509,PhysRevB.80.214514}. These two components are further shown not to be isolated, but instead entangled~\cite{2011arXiv1103.5073Z}.

\section{Substitution effects of 3$d$ transition metals \label{sec:subeffect}}
There have been extensive studies on the effects of substitution with 3$d$ transition metals (such as Co, Ni, Cu, etc) in the Fe-based superconductors~\cite{canfield:060501,PhysRevLett.110.107007,PhysRevLett.109.077001,PhysRevB.84.020509}. Some initial studies on BaFe$_2$As$_2$ (Ba122) have suggested that 3$d$ metals such as Co, when partially substituted for Fe, act as effective electron donors~\cite{canfield:060501,PhysRevB.82.024519,PhysRevB.84.020509}; this process can be described using a rigid-band shift model~\cite{PhysRevB.83.094522,PhysRevB.83.144512,JPSJ.80.123701}.  However, such a model has faced serious challenges from both experimental and theoretical perspectives~\cite{PhysRevLett.105.157004,PhysRevLett.110.107007,2011arXiv1112.4858B,2011arXiv1107.0962B,PhysRevLett.109.077001}. Furthermore, there is a dichotomy between Co/Ni and Cu substitution effects~\cite{canfield:060501,PhysRevB.82.024519,PhysRevB.84.020509,PhysRevB.84.054540,2011arXiv1112.4858B,0953-8984-24-21-215501,PhysRevLett.110.107007}. 
In \fts, Cu is also found to have a stronger suppression on both the conductivity and superconductivity than Co and Ni~\cite{danielfeseco,williams-2009-21,PhysRevB.82.104502,mkwreview1,JPSJ.78.074712,2010arXiv1010.4217G,fetesenico1,Zhang20091958}. 

The author and collaborators have performed resistivity and inelastic neutron scattering measurements on three samples of Fe$_{0.98-z}$Cu$_z$Te$_{0.5}$Se$_{0.5}$ with $z=0$, 0.02, and 0.1 (labelled Cu0, Cu02, and Cu10 respectively)~\cite{PhysRevB.88.144509}.  It is found that with increasing Cu doping the sample's resistivity deviates progressively from that of a metal. The Cu02 (in the normal state) and Cu10 samples behave like three-dimensional Mott insulators, similar to the behavior in Cu-doped FeSe for Cu doping larger than 4\%~\cite{PhysRevB.82.104502,williams-2009-21}. Meanwhile, the low-energy ($\leq 12$~meV) magnetic scattering is enhanced in strength, with greater spectral weight and longer dynamical spin-spin correlation lengths. Such an enhancement of the low-energy magnetic scatterings by Cu doping is intriguing. One plausible interpretation of the results is that the main effect of Cu substitution is to introduce localization into the system and suppress the itinerancy, and thus to enhance the magnetic correlations~\cite{localm3,0953-8984-24-21-215501}. In the case of Cu-doped FeSe, it has also been suggested that Cu substitution introduces local moments, and when Cu doping equals 0.12, the sample exhibits a spin-glass transition~\cite{williams-2009-21}. More recently, our group have also found evidence that Co/Ni and Cu have different effects on magnetic excitations~\cite{PhysRevB.91.014501}, consistent with the different responses of transport properties to these impurities~\cite{danielfeseco,williams-2009-21,PhysRevB.82.104502,JPSJ.78.074712,2010arXiv1010.4217G,fetesenico1,Zhang20091958}.

The effects of Cu substitution in the parent compound Fe$_{1.1}$Te have also been studied~\cite{arXiv:1205.5069}. It is found that Cu drives down the structural and magnetic transitions, with long-range nearly-commensurate magnetic order retained in Fe$_{1.06}$Cu$_{0.04}$Te, but only short-range incommensurate order in FeCu$_{0.1}$Te~\cite{arXiv:1205.5069}. In the latter sample, the structural phase transition is not obvious and a transition to a spin-glass state is found at 22~K~\cite{arXiv:1205.5069}. In Fe$_{1.06}$Cu$_{0.04}$Te, the initial structural and magnetic ordering occurs at 41~K, involving short-range incommensurate order that abruptly shifts to long-range nearly-commensurate order below 36~K~\cite{arXiv:1205.5069}. Inelastic scattering measurements indicate a spin anisotropy gap of 4.5~meV in the nearly-commensurate phase~\cite{arXiv:1205.5069}. The results are consistent with the idea that the frustration of the exchange interactions between the coupled Fe spins increases as more Cu is added. Although both the structural phase transition and magnetic order are suppressed with increasing Cu substitution, superconductivity is not observed~\cite{arXiv:1205.5069}. The key issue is likely associated with the spin configurations. Although there is a suppression of the magnetic order at (0.5,\,0), no evidence of the enhancement of the C-type spin correlations has been observed. This is another piece of evidence for the importance of low-energy fluctuations near (0.5,\,0.5) to the superconductivity in \fts.

\section{Summary}
\label{sec:conclusion}
To summarize, I review the progress of neutron scattering studies on the Fe-based superconductor \fts. Many important results concerning the relationship between spin and superconductivity have been reported thus far. Particularly, while the bicollinear spin correlations appear to compete with superconductivity, the collinear correlations seem to promote it. I have also shown that both local moments and itinerant electrons may contribute to magnetic excitations. Substitution with different 3$d$ transition metals may have varying effects in the 11 system; for example, the main effect of Cu is to induce localization.

Although the research pace has been extremely fast, there still remain many challenges. For example, it appears relatively easy to make large single crystals for \fts ($x\leq0.7$), but it is difficult to have them be homogeneous~\cite{PhysRevB.82.020502,phasese11,PhysRevB.83.220502}. Also, the extra Fe can sometimes complicate interpretation. Therefore, homogeneous samples with more precisely controlled stoichiometry may provide a better understanding of the physics. This goal may be achieved by using a smaller cooling rate during the growth, or annealing the as-grown crystals~\cite{taen:092502,JPSJ.79.084711,yeh-2009,revisedfetesephase,2011JAP93914M}.  Given the many interesting reports on the thin-film samples of FeSe~\cite{sr4_6040,2012arxiv1201.5694W,nm_12_634,nm_12_605,arXiv:1202.5849,arXiv:1406.3435}, it is certainly worth putting more effort to growing single crystals of FeSe. Recent studies have reported the growth of phase pure crystals of superconducting FeSe~\cite{Hu2011,PhysRevB.90.144516}. On the measurement side, there are still many things that can be done. For example, the origin of the resonance and the magnetism are still not fully understood, and require further investigation.

\section{Acknowledgements}
The work at Nanjing University was supported by National Natural Science Foundation of China under Contract No. 11374143, Ministry of Education under Contract No. NCET-13-0282, and Fundamental Research Funds for the Central Universities. I thank all of the collaborators listed in the references, especially those at Brookhaven National Laboratory and University of California at Berkeley, where I had worked; and the colleagues in the Physics Department of Nanjing University, where I am now affiliated. I am grateful for the colleagues and collaborators for allowing their work to be reproduced here. I thank J. A. Schneeloch for proof reading.

\end{document}